# A Space-Time Knife-Edge In Epsilon-Near-Zero Films for Ultrafast Pulse Characterization

Adam Ball[1]*, Ray Secondo[2,3], Dhruv Fomra[1,4,5], Jingwei Wu[1,6], Samprity Saha[1], Amit Agrawal[5], Henri Lezec[5] Nathaniel Kinsey[1]*

[1] Virginia Commonwealth University Department of Electrical & Computer Engineering Richmond, VA 23284
[2] Azimuth Corporation, Beavercreek, OH, 45324
[3] Materials and Manufacturing Directorate, Air Force Research Laboratory, Wright-Patterson Air Force Base, OH, 45433
[4] Department of Chemistry and Biochemistry, University of Maryland, College Park, MD 20742
[5] Physical Measurement Laboratory, National Institute of Standards and Technology, Gaithersburg, MD 20899
[6] College of Optical Science, University of Arizona, Tucson, AZ 85721
*Corresponding authors: ballar@vcu.edu, nkinsey@vcu.edu

**Abstract (200 words)**

Epsilon-near-zero (ENZ) materials have shown strong refractive nonlinearities that can be fast in an absolute sense. While continuing to advance fundamental science, such as time varying interactions, the community is still searching for an application that can effectively make use of the strong index modulation offered. Here we combine the effect of strong space-time index modulation in ENZ materials with the beam deflection technique to introduce a new approach to optical pulse characterization that we term a space-time knife edge. We show that in this approach, we are able to extract temporal and spatial information of a Gaussian beam with only two time resolved measurements. The approach achieves this without phase-matching requirements (<1 μm thick film) and can achieve a high signal to noise ratio by combining the system with lock-in detection, facilitating the measurement of weak refractive index changes ($\Delta n \approx 10^{-5}$) for low intensity beams. Thus, the space-time knife edge can offer a new avenue for ultrafast light measurement and demonstrates a use cases of ENZ materials. In support of this, we outline temporal dynamics for refractive index changes in non-colinear experiments opening avenues for better theoretical understanding of both the spatial and temporal dynamics of emerging ENZ films.

## Introduction

Recently, the epsilon-near-zero (ENZ) regime of materials has received significant attention from the photonics community due to its ability to achieve intriguing optical effects [1–5], among which the enhancement of nonlinear optical responses is one of the most celebrated [6–9]. ENZ based nonlinear enhancements have been demonstrated in thin films (metals, doped semiconductors, and phononic materials) as well as in structured systems (see Ref. [10,11] for other ENZ materials). However, the family of transparent conducting oxides (TCOs) (e.g. doped ZnO, CdO, $Sn_2O_3$) have seen significant attention due to their ability to achieve an ENZ region in the near-infrared spectral range while exhibiting reasonably low loss ($\varepsilon'' \approx 0.3$), broadband and tunable properties, being widely available, and with well-established CMOS compatibile deposition techniques [12–15].

Numerous works have shown enhancement to processes including harmonic generation [16–23], wave-mixing [24,25], negative refraction [26,27], and index modulation [28–32]. Among these, one of the most intriguing features of TCOs is their ability to achieve near-unity index modulation on the picosecond timescale across the entire near-infrared spectral range. The origins of the strong index tuning in TCOs arise from the unique combination of slow-light enhancement provided by the dispersive ENZ region ($n_g \sim 4\text{-}10$), the ability to employ free-carrier nonlinearities that are still *fast enough* (picosecond scale), and a high damage threshold (>1 TWcm$^{-2}$ in the infrared) [16,28]. Placing this into context, we can compare this to III-V quantum well structures, one of the leading nonlinear materials, and note that the typical index change observed is $\Delta n_{QW} \approx 0.14$ with $\alpha \approx 1000$ cm$^{-1}$ giving rise to a $FoM = \frac{1}{L_\pi \alpha} \approx 0.7$. TCOs at ENZ have shown the ability to achieve $\Delta n_{ENZ} \approx 0.72$ with $\alpha = 27{,}362$ cm$^{-1}$ and $FoM \approx 0.4$, in other words, a $\pi$ phase shift in less than 1 µm with approximately 90% transmission [6,33]. Moreover, ENZ can achieve this in a single thin film without complex growth or nanofabrication and does so with a bandwidth of 300 nm or more. Thus, it is apparent that ENZ materials are among some of the most promising nonlinear materials available.

The combination of such fast and substantial index tuning has inspired significant research to understand the extreme effects in ENZ materials, including tunable cavities [34,35], bi-stability [35,36], plasmon switching [37,38], as well as frequency shifting and other time-varying interactions [39–45]. Despite this excitement and the unique capabilities offered by ENZ films, their impact on applications of interest has remained an unanswered question primarily due to the loss, which creates a performance tradeoff. For in-plane geometries (e.g. on-chip), the loss is mitigated by reducing mode coupling with the ENZ material [46,47], while in free space it is minimized by employing films of sub-micron thickness. Along these lines, recent efforts have shown that ENZ films can provide improved THz emission [48,49] through ps-scale modulation of the conductivity, and compact frequency-resolved optical gating pulse characterization due to the strong harmonic generation in the thin film [50]. However, the use of the strong optical refractive nonlinearity, a salient and oft-praised feature of ENZ, has yet to be exploited in an application scenario.

Here we fill this gap by leveraging the strong and fast refractive nonlinearity of a sub-micron-thick TCO film at ENZ to achieve a 'space-time knife edge' for ultrafast pulse characterization. By combining the space-time knife edge with the beam deflection geometry [51–55], which intentionally displaces the centroid of the pump and probe beams, we demonstrate that the approach generates unique cross-correlation signals that depend upon the spatial and temporal properties of the beam. The technique thus allows for the extraction of important beam parameters such as spatial width, centroid deviation, temporal width, angle of arrival, and phase front tilt down to the femtosecond and micron scales while requiring only two measurements on a segmented detector. It thus constitutes an improvement upon the widespread intensity autocorrelation (IA) technique, capable of resolving spatial parameters without the use of second harmonic generation (SHG) or a camera. Space-time knife edge expands the spectral and rate limitations of IA, useful for rapid and simplistic characterization of pulses in research labs. Furthermore, by using a quadrant-cell detector rather than a camera, we can utilize a lock-in amplifier to measure extremely small signals for a high signal to noise ratio. Using $\chi^{(3)}$ enables us to expand our sensitivity down to low intensity levels ($\sim 500$ MWcm$^{-2}$) while still employing a very thin film (~100's of nm).

Additionally our efforts highlight the consequences of strongly noncolinear pulses interacting in thin ENZ media. This often used configuration allows one to exploit normal electric field confinement in the ENZ layer, but as we show, modifies the temporal dynamics. While this effects provides a rich interaction, accounting for these effects is critical for accurate interpretation nonlinear measurements such as reflection and transmission modulation, beam deflection, Z-scan, frequency shifting, and others.

**Results and Discussion**

*Ultrafast Pulse Characterization*

Optical pulse characterization is of fundamental importance in ultrafast science, where determining properties such as spatial width, temporal width, polarization, frequency spectrum, spectral phase, and others is a common practice in many labs. Many techniques to accomplish this exist including intensity autocorrelation and frequency-resolved optical gating (FROG), see Refs. [56,57] and [58] for more information and tutorials on pulse characterization. Among the available techniques, IA is one of the most common techniques employed to determine the pulse width of stable sources (such as Ti:sapphire and fiber lasers) due to its simple structure, robustness, and low cost. For fully unknown or unreliable sources, more detailed characterization techniques such as FROG or SPIDER are suggested to avoid the coherent artifact [59–61], see more in the appendix [62,63].

In commercial systems, IA is performed by evenly splitting a beam while recombining the daughter beams within a second-order nonlinear material to generate a second-harmonic signal. By varying the path length of one of the arms, information on the temporal shape of the pulse is obtained through the intensity of the harmonic generation. While simple, IA requires the use of bulk nonlinear crystals to achieve sufficient signal amplitude that must be properly oriented to achieve phase matching for harmonic generation. Additionally, in the conventional configuration, any angular offset between the beams is ignored as the measurement consists only of the total transmission intensity. As such it is sensitive to temporal variations, but not sensitive to spatial properties of the beam and must be supplemented by other tools to provide additional characterization.

Here we seek to overcome this limitation by expanding the functionality of IA to enable the characterization of spatial widths, pulse front tilt, and centroid displacement using a refractive nonlinearity in a thin ENZ film combined with a spatially sensitive measurement to completely remove phase-matching constraints while maintaining strong signal strength.

*The Space-Time Knife Edge in ENZ Films*

The core of the effort focuses on exploiting the interaction of a probe pulse with the space-time knife edge generated through off-axis excitation by a strong pump beam. In the coordinate system of the sample, the irradiance of a Gaussian excitation beam can be described by Eq.(1):

$$I_{pump}(x,y,t) = n_g I_0 e^{\left(-(a(x-x_o)^2 + 2b(x-x_o)(y-y_o) + c(y-y_o)^2)\right)} e^{\frac{-(t-t_0)^2}{2\tau_{pu}^2}}, \quad (1a)$$

$$I_{probe}(x,y,t) = I_0 e^{\left(-\frac{x^2}{2w_x^2} - \frac{y^2}{2w_y^2}\right)} e^{\frac{-(t)^2}{2\tau_{pr}^2}} \quad (1b)$$

where $n_g$ is the group index, and $\tau_p$ is the 1/e² temporal width. Variables $x_o, y_o$ are offsets describing the position of the pump beam. Coefficients a, b, and c describe the rotation of a Gaussian beam in space and contain $\theta$, the angle of incidence, $w_x$ and $w_y$, the 1/e² spatial widths described in the appendix. For nonlinearities with a finite relaxation time $\tau_r$, the index perturbation is determined by solving the rate equation to find the energy density $U$ deposited into a thin sample in space and time, which is directly proportional to the change in index [30,32]:

$$U(x,y,t) = e^{\frac{-t_{diff}}{\tau_R}} \left[ \frac{A \cdot \left(erf\left(\frac{2v\tau_R C - 1}{B}\right) - 1\right)}{B} \right] \quad (2)$$

where $A = -\sqrt{\pi} e^{-\frac{4v\tau_R x(acv\tau_R x - b^2 v\tau_R x + b + a) - 1}{B^2 \tau_R^2}}$, $B = 2\sqrt{c + 2b + av}$, and $C = bx + ax - ctv - 2btv - atv$, where $v$ is the velocity of the wavefront outside the material, see appendix for derivation. In this configuration, the finite angle of incidence produces a short time delay ($t_{diff} = \frac{w_x}{v}\sin(\theta)$) between the arrival of one side of the beam and the other, as illustrated in Fig. 1a.

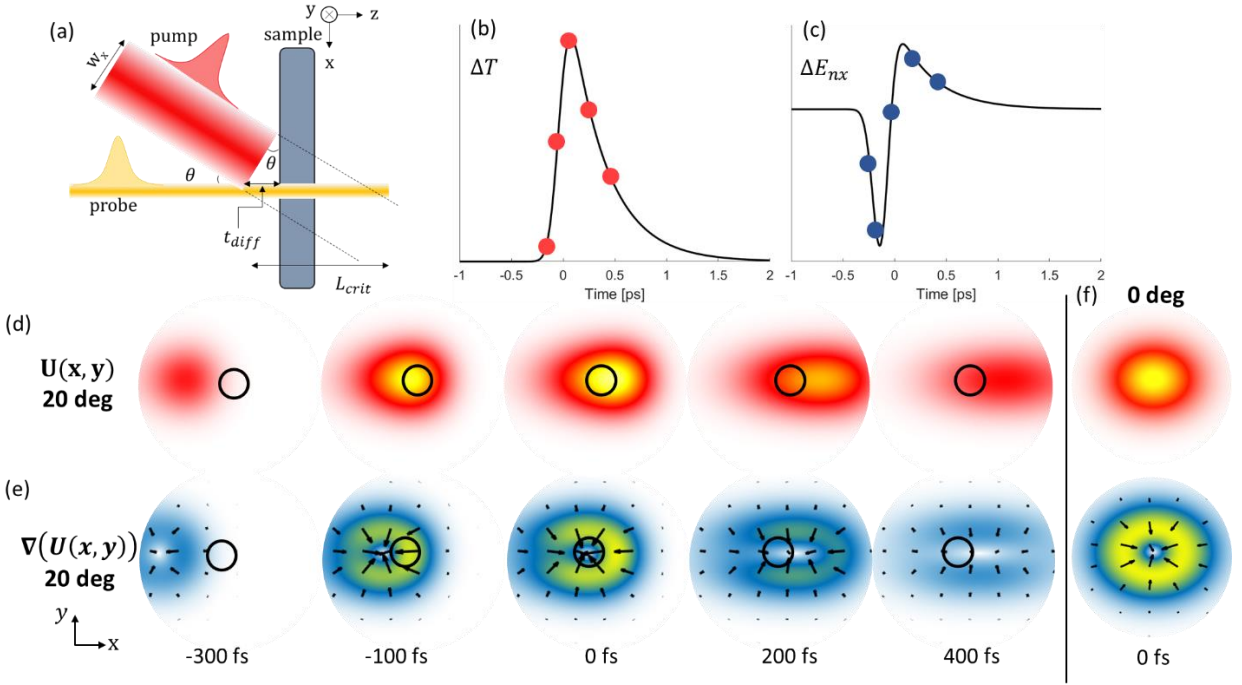

Figure 1: (a) XZ-plane of the pump beam arriving on a sample at an angle. The difference in time from the arrival of one side versus the other is denoted by $t_{diff}$. $L_{crit}$ defines the critical distance at which the two beams no longer overlapped. (b) Change in transmission ($\propto U(x,y)$) of the probe at versus time at probing position indicated by the black circle in (d). (c) Change in deflection of the probe ($\propto \nabla U(x,y)$) versus time at probing position indicated by the black circle in (e). (d) XY-planes at sample surface illustrating energy distribution U(x,y) at five times of -300, -100, 0, 200, and 400 fs delay between pump and probe beams. (e) XY-planes at sample surface illustrating the slope of energy distribution $\partial U(x,y)$ at five times of -300, -100, 0, 200, and 400 fs delay between pump and probe beams. (f) Normal incidence pump beam illustrating at all times U(x,y) and $\nabla U(x,y)$ is centered because there is no angle-induced gradient present.

Thus, for a pump beam incident at an angle $\theta$, there is a *shape-induced* refractive index change created by the projection of the excitation intensity into the plane of the sample at any given moment (timesteps of Fig. 1d), as well as an *angle-induced* refractive index change created from the sweeping motion of the beam colliding with the sample surface at different times as illustrated by a movement of the center of mass of $U(x,y)$. This gives rise to a so-called 'photonic guillotine' [64], that implicitly contains information about both the spatial and temporal properties of the excitation beam, which is made ultrathin here to for a 'space-time knife edge'.

For ENZ films, $U(x,y,t)$ is directly proportional to the change in transmission of the film $T(t)$, Fig. 1b [10,65,66]. However, as is evident in Eq. (2), the space- and angle-induced effects are convolved in $U(x,y)$ and can be difficult to distinguish with a single measurement of transmission. To separate them we employ a modified beam deflection technique, see Appendix for full details. In this measurement, a strong pump beam modulates the ENZ material's refractive index via effective third-order nonlinearities (as described in [10,65,66]) and a weaker, smaller probe beam (typically 3-5× smaller) is placed on the shoulder of the pump beam in the region of quasi-linear intensity. During modulation, the probe beam is physically deflected due to the local refractive index gradient causing a spatial shift of the position ($\delta x$ and $\delta y$) that is measured by a quadrant cell detector placed a distance $L$ away. The gradient of absorbed energy, $\nabla U(x,y)$, also experiences a shape-induced and angle-induced effect, shown in Fig. 1c and 1e. However, unlike transmission, the deflection signal is sensitive to direction. We can quantify the normalized deflection $\Delta E_n$ as:

$$\delta x = \frac{L n_{2,eff} I_{pu} t}{w_{pr}} \propto \frac{\partial U(x,y,t)}{\partial x} \tag{3a}$$

$$\delta y = \frac{L n_{2,eff} I_{pu} t}{w_{pr}} \propto \frac{\partial U(x,y,t)}{\partial y} \tag{3b}$$

$$\Delta E_{n,x}(t) = \frac{E_{left} - E_{right}}{E_{total}} \propto \frac{\left[\iint_{Left} I_{pr}(x-\delta x, y, t) dA - \iint_{Right} I_{pr}(x-\delta x, y, t) dA\right]}{\iint_{A_{det}} I_{pr}(x-\delta x, y, t) dA}, \tag{3c}$$

where $n_{2,eff}$ is the effective change in refractive index for the ENZ film, $L$ is the distance from sample to detector, t is the thickness of the film, and $A_{det}$ is the detector area. For deflection along y ($\Delta E_{n,y}$), one would simply integrate over the top and bottom half of the detector and use the spatial shift $\delta y$ instead of $\delta x$.

Notably, we can see that for a beam incident in the x-z plane as shown in Fig. 1a, the angle-induced index change only contributes to the gradient along the x-direction, see Eq. 3. This is also evident in Fig. 1e by the movement of the pump center of mass (depleted center) across the probe, which generates a bipolar deflection signal as in Fig. 2c. This ensures that every point within the pump experiences an x-direction deflection at some point in time during the interaction. This is not present at normal incidence, Fig. 1f, where x-direction deflection is zero directly over the center of mass of the absorbed energy density. Conversely, the distribution of the y-gradient remains fixed in space (similar to normal incidence), with magnitude rising and falling with temporal shape of the material response. Thus, a measurement of y-deflection quantifies the shape-induced effect and can be used to normalize the x-deflection which contains both signals.

*Measuring the Space-Time Gradient*

Using this configuration, subsequent measurements were done using cross-polarized non-degenerate beams with an excitation beam at 1400 nm and a peak irradiance of 56 GWcm$^{-2}$, and a probe beam at 1300 nm (both beams $\Delta\lambda_{FWHM}$ = 35 nm). These pulses were generated from an 800 nm Ti:Sapphire laser with a pulse width of 97 fs. For more details see Methods. For a pump and probe beam incident in the x-z plane with the pump arriving at an angle θ from normal, we measure $\Delta E_n$ in both the x- and y-directions for the locations P1-P5 on the beam, see inset in Fig. 2, by introducing an offset $\Delta x$ (relative to the center of the excitation) which moves the beam to positions P1, P2, or P3 respectively. For positions P4 and P5, the beam is moved vertically by an offset of $\Delta y$ relative to the center of the pump beam at time zero.

Let us first examine the deflection signal in the y-direction in Fig. 2. Here we note only P4 and P5 observe a strong y-direction deflection as expected. Since the near infrared excitation excites the intraband nonlinearity in the ENZ film, the index tends to increase with intensity [30]. In this case the beam deflects towards the region of higher index resulting in positive $\Delta E_y$ at P4 and a negative $\Delta E_y$ for P5. For P1, P2, and P3, the y-direction deflection is minimal as $\partial U(x, y_o)/\partial y \sim 0$ on the horizontal axis of the Gaussian-shaped pump, agreeing with Fig. 1e

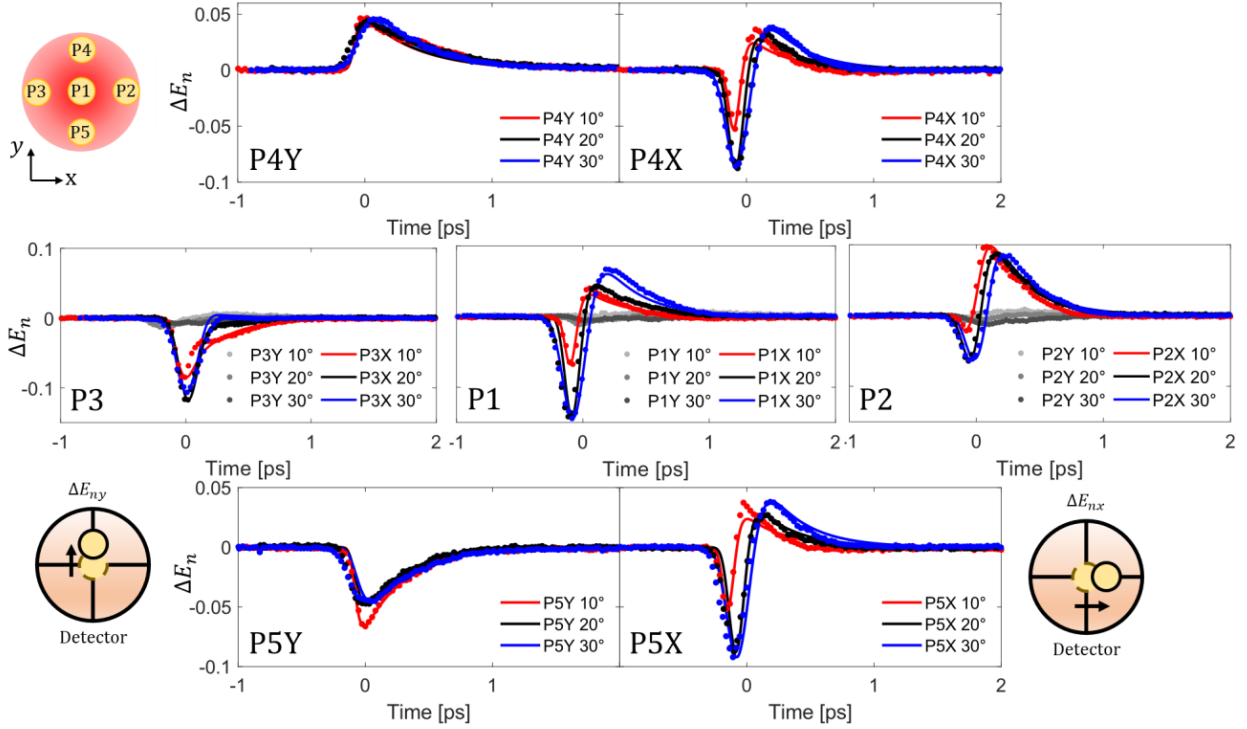

Figure 2: $\Delta E_n$ in the X and Y-directions on the quadrant cell for points P1-P5 at all three angles of 10°, 20°, and 30°. There are only two Y-direction $\Delta E_n$ signals P4Y and P5Y which provide significant signal due to the spatial gradient, the others are illustrated by grey dots denoted by the legends.

Next let us examine the results of the deflection in the x-direction. As expected from Fig. 1e, we see that due to the angle-induced effect modifying the index across the surface of the sample at different times, all positions on the sample experience a deflection. As this gradient occurs on the excitation (t < 0) and relaxation (t > 0) of the index perturbation pulse, the corresponding trace exhibits a bipolar response. For ultrafast nonlinearities, the angle-induced deflection signal would be temporally symmetric (see supplemental for results on fused silica), but is asymmetric here due to the temporal relaxation arising from the ENZ response function (Fig. 2 P4Y).

It is also worth noting a few additional subtleties to the angle-induced gradient signal that can be used to provide additional verification of the extracted values. First, at positions P1, P2, or P3, the arrival time of the pump excitation relative to the probe is determined by the value of $t_{diff}$ in Eq.(3), thus resulting in an apparent delay of the signal. This delay is directly linked to angle of incidence, phase front tilt, width of the excitation, and position of the probing beam relative to the excitation. Secondly, as the angle of incidence increases, the angle-induced deflection has a larger impact on $\Delta E_{n,x}$ because the rising edge of the energy becomes steeper in space and produces a more asymmetric cross-section of the pump (compare Fig. 2 angles 10°, 20° and 30°). Conversely, smaller angles results in a quenching of the angle-induced deflection as all regions of the sample are excited almost simultaneously. Performing this experiment with colinear beams is expected to eliminate the angle-induced gradient, however, performing degenerate colinear excitation is not easy due to the challenge of separating the excitation from the probe on the detector as well as increased influence of parasitic effects such as two beam coupling [67,68].

Third, the ratio between the temporal and spatial width of the excitation beam dictates how quickly or slowly the bipolar response is established. If the widths are asymmetrical in space ($w_x$ versus $v\tau_p$), the temporal spacing between the peak and valley of $\Delta E_n$ is condensed, and is elongated for more symmetrical pulses. With these additional verifications, one could either directly calculate the excitation parameters or fit the magnitudes of the unique peaks from Eq. 2. This provides a unique solution to the beam deflection signal given the relaxation and pulse width is extracted initially from P1 and P4.

*Measuring an Unknown Pulse via Space-Time Knife Edge*

With fundamentals of the measurement outlined above, we illustrate an example measurement schematic employing the space-time knife edge (STKE) technique. Under the assumption of a radially symmetric beam, only a subset of the data from P1 – P5 is needed.

First, we split the excitation beam by picking off two weaker identical daughter beams one of which is spatially placed at the center of the excitation (P1), and one which is vertically offset by $(1/e)w_e$ (P4), or peak $\Delta E_{n,y}/E$. Next, we sweep in time the pump beam with respect to the excitation (using a delay stage or vibrating mirror), resulting in the two plots in Fig. 3.

The measurement of y-direction deflection at positions P4 (Fig. 3a) allows us to obtain a measurement combining the pump and probe cross correlation and intrinsic material relaxation $\tau_R$. Under the assumption that the pump and probe are temporally identical, this curve can be fit to Eq. 2 to extract the pulse width $\tau_p$, the relaxation of the material $\tau_R$, and the refractive index change $\Delta n$ due to the spatial gradient only. Next, the measurement of the x-direction deflection at P1 (Fig. 3b) can be fit to Eq. 2 to extract the angle of arrival or phase front tilt of the excitation beam $(\theta, \alpha)$ [69]. Additionally, each point provides an implicit transmission measurement $\Delta T/T = \Delta E_{total}/E_{total}$ which captures the local excitation intensity. By utilizing the change in transmission value at P1 as well as the values of the index gradient at P4 or P5, the width of the Gaussian excitation pulse $w_y$ can be extracted. This is done by setting the peak of the Gaussian to the maximum $\Delta T$ value measured at P1. The $\Delta T$ value retrieved at P4/P5 will be at the maximum slope of the Gaussian, or the 1/e half-width, thus directly extracting $w_y$.

Overall, this configuration gives us a unique ability to separate both refractive index gradients and extract spatial and temporal information of the pump. For example, by utilizing Eq.(2) we can fit the experimental results and extract $\tau_p \approx 95 \pm 5$ fs, $\tau_R \approx 400 \pm 20$ fs, and $w_x \approx 600 \pm 30$ μm. The result fits well to all measured values and is consistent for multiple angles of incidence (10°, 20°, and 30°) as well as for multiple materials including Al:ZnO and Fused Silica (see supplemental).

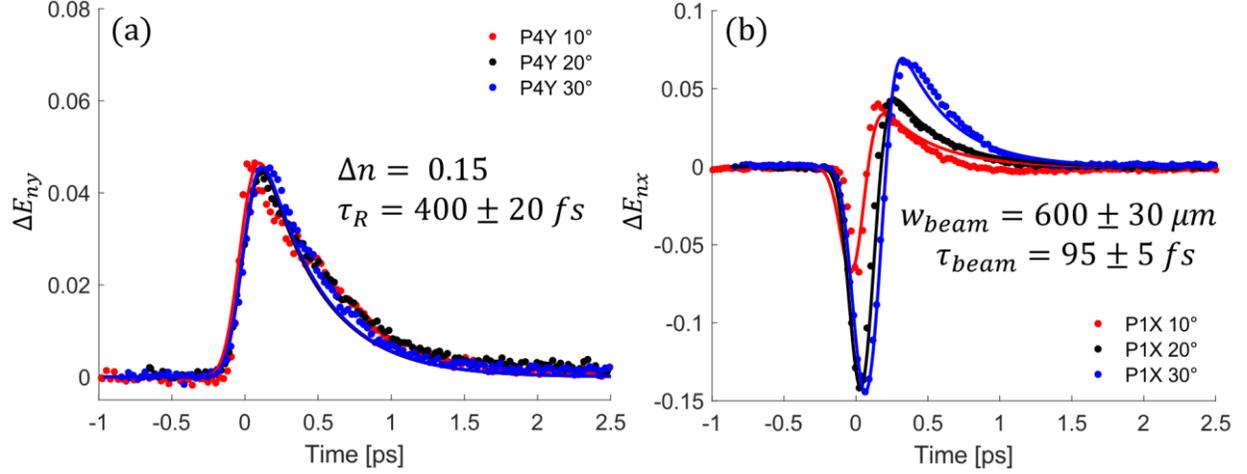

Figure 3: (a) Spatial intensity gradient to calculate relaxation and changes in $\Delta n$ at P4Y. (b) Angular-induced gradient to calculate the beam spatial and temporal width at P1X based on the modeling of (a).

If the beam is not spatially symmetric, either P2 or P3 can also be measured by placing the beam at the 1/e point in the horizontal direction and recording the $\Delta E_{n,x}$ signal. Moreover, if we wish to fully characterize the pump beam, a systematic sweep of the probe position across the pump can be conducted. Utilizing the gradient equation, precise determination of the spatial relationship between the two beams on the micron scale becomes achievable by analysis of both the temporal delay of each signal and the relative magnitude of the positive and negative peaks. An example of this at 10° can be seen below in Fig. 4a and 4b. When the relative beam position between the pump and probe are offset by 30 μm (blue versus purple line in Fig. 4a) a 3.5% $\Delta E_{n,x}$ signal can be distinguished. Given that our system is capable of measuring 0.1% $\Delta E_{n,x}$, a spatial resolution of ~1 μm is achieveable. This is comparable CMOS beam imagers with pixel sizes on the order of 1-2 μm and an order of magnitude smaller the near infrared imagers.

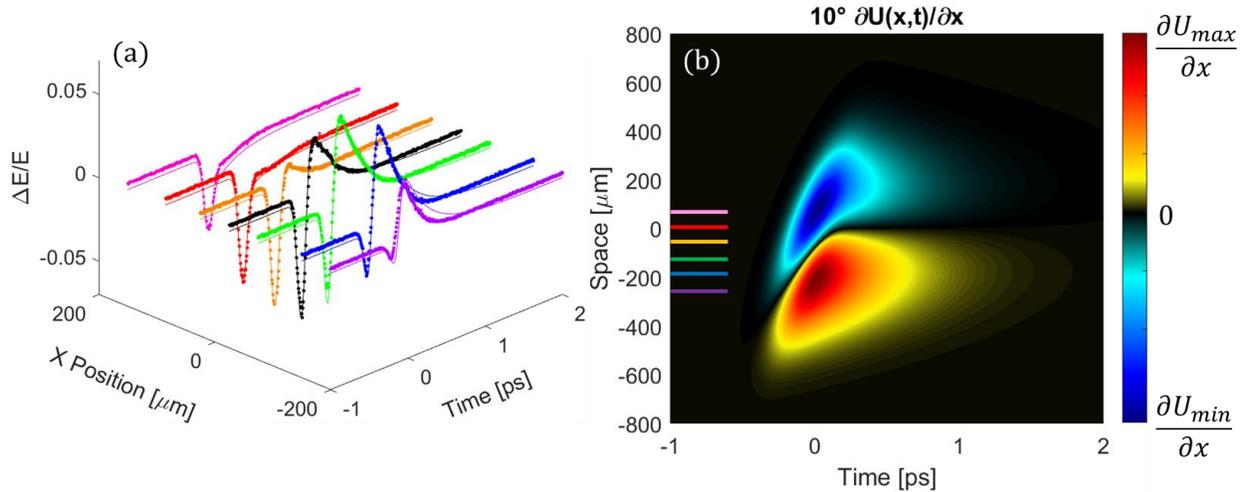

Figure 4: (a) Slopes along excitation direction for various probe beam positions. The relative heights between positive and negative peaks allows for one to back calculate the position of the two beams. (b) Surface plot of the slope of the energy distribution and relative probe positions illustrated by horizontally colored lines representing position of the probe beam.

## Discussion

Fundamentally, the approach utilizing the space-time knife edge can be viewed as an improved version of intensity autocorrelation. As a result it is important to compare the two and discuss the differences. For a second harmonic process, the IA trace follows equation 4 below.

$$A_{IA}(\tau) \propto \int_{A_{det}} \int_{-\infty}^{\infty} \eta I(t) I(t-\tau) dt dA \qquad (4)$$

Observing Eqs.(3), we can see that the major difference between the output of the modified beam deflection setup to the normal IA setup is the spatial and directional sensitivity that provides further degrees of freedom to measure changes in autocorrelation at a mixture of space and time steps. This allows thorough characterization of the input spatial width, angle, centroid deviation, and phase front tilt. Consequently, providing more information obtained than the conventional IA.

To achieve this added spatial sensitivity, the STKE setup utilizes ENZ materials while in the IA setup SHG crystals are preferred. This simplification eliminates the need for phase matching, and ENZ materials can offer broader bandwidth compared to SHG crystals. Notably, for pulses 10 fs and below, IA is difficult due to the coherent artifact and phase matching constraints [70]. For STKE, pulse width limitations reside in temporal and bandwidth limits of the material response where the index change in driven redistribution of the electron population [30]. The first restriction on the framework outlined here is the thermalization time. For pulses less than or equal to the electron-electron collision time, the distribution of the electron population is athermal. In this case, STKE is still feasible, but additional temporal signatures may arise which would require an expanded analysis. Presently, there is limited data on athermal processes for ENZ films, but it is assumed that the electron-electron collision time is ~10 fs leading to thermalization of the population on the scale of ~30 fs. Y. Sivan et al. have provided this analysis for metals, but for ENZ materials this is still an open question [71]. The second restriction occurs due to the absorption enhancement region of the material, which drives the index change, and the region of strong refractive index dispersion, which results in deflection. These conditions create an upper and lower bound upon the bandwidth which is equivalent to is the nonlinear enhancement bandwidth $n_g^2$, shown to be ~300 nm for most commonly explored oxides. Assuming a transform limited pulse with a 300 nm bandwidth and a 1500 nm center wavelength, this provides 8-10 fs, which would represent limit of technique and is comparable to the limts outlined for ENZ-FROG [50]. However, we note that using an instantaneous polarization-driven nonlinearity removes these pulse restrictions, although the nonlinearity is proportionately 100x weaker and a higher irradiance would be required, see results of fused silica in the supplemental.

Furthermore, because we are interested only in a 1-dimensional signal, a lock-in amplifier can be used to enhance the signal to noise ratio. This allows us to detect extremely small signals with beam deflection energy displacement ($\Delta E_n$) on the order of $1 \times 10^{-3}$ ($\theta \approx 1$ mrad) correlating with refractive index perturbations ($\Delta n$) on the order of $10^{-5}$ and intensities on the order of 500 MW/cm² [51]. Such capabilities allow for the $\chi_{eff}^{(3)}$ in thin ENZ films to be on par with materials utilizing bulk $\chi^{(2)}$ crystals and avoids the need for long interaction lengths (mm scale) where beam walk-off within the sample ($L_{crit} = \frac{w_{beam}}{\sin(\theta)}$ Fig. 2b), limits the performance.

Although the approach presents advantages, it is also important to understand its limitations. First, the STKE technique (like conventional IA) assumes a symmetrical beam shape in space and time, typically Gaussian, and requires a probe beam(s) that is 3-5x smaller than the excitation. The relative variation in sizes places limits on the range of input beam sizes that could be measured by a single system, as internal optics to shrink the beam would be needed and their performance would vary based on the input size. Additionally, because the technique employs a non-instantaneous nonlinearity, phase information (with the exception of the phase front tilt) is implicitly lost, although it could be recovered by employing Kerr nonlinear materials such as fused silica at the price of requiring increased interation length or irradiance. Further development with ENZ materials could realize alternatives to extract phase information through the use of interference effects [72] employing techniques such as two-beam coupling at the surface of the sample.

Lastly, we wish to highlight the impact of the angle-induced gradient on nonlinear experiments in off-axis excitation, as is popular to do with ENZ materials due to the off-axis boundary condition enhancement [6,40,73,74]. As shown in Fig. 2, the increasing angle results in an asymmetric beam at the surface. If one were to do a Z-scan or reflection/transmission measurement at an angle, the space-time characteristics of the beam are not captured. This angular separation can result in asymmetric focusing due to a stronger pump gradient on the rising edge than falling edge which deviates from the expected response for normal incidence and can dominate the response at large angles (e.g. >30º). Therefore, exercising caution is essential when undertaking nonlinear experiments with non-colinear or angled setups.

To conclude, the space-time knife edge technique is a unique opportunity to join the temporal resolution of femtosecond autocorrelation and the spatially resolved information of beam deflection. This technique is unique in the fact that it makes practical use of strong nonlinear refractive provided by ENZ materials – perhaps their most salient feature - which directly facilitates the use of extremely thin films.

**Acknowledgements**

Acknowledgments. Dr. Nathaniel Kinsey acknowledges the Air Force Office of Scientific Research (Nos. FA9550-1-18-0151 and FA9550-16-10362). Dr. Henri Lezec and Dr. Dhruv Fomra also acknowledge support under the Professional Research Experience Program (PREP), administered through the Department of Chemistry and Biochemistry, University of Maryland. Research performed in part at the National Institute of Standards and Technology Center for Nanoscale Science and Technology. The views and conclusions are those of the authors and should not be interpreted as representing the official policies of the funding agency. Certain equipment, instruments, software, or materials, commercial or non-commercial, are identified in this paper in order to specify the experimental procedure adequately. Such identification is not intended to imply recommendation or endorsement of any product or service by NIST, nor is it intended to imply that the materials or equipment identified are necessarily the best available for the purpose. We wish to acknowledge the Air Force Office of Scientific Research for funding this work under Grant No. FA9550-22-1-0383.

SUPPLEMENTAL:

*Pulse Characterization and Limitations*

Temporal pulse characterization methods began with the idea of interferometry. Two separate fields interfere and produce a unique pattern dependent upon spatial, temporal, and or spectral overlap within some reference sample. Typically, one optical beam is a reference beam which is well characterized beforehand, or the two beams stem from one source which has been split and thus the recovered beam is a self-referenced result. In any case, one can think about all of the particular characteristics that an optical researcher may want to know about an optical beam including (not limited to) spatial width, temporal width, polarization, phase, and frequency spectrum, resulting in the need for multiple measurement devices for any combination of the listed features. The most popular techniques for ultrafast pulses can be separated into categories of spectrally resolved (FROG, SPIDER), intensity resolved (autocorrelation, TIGER), or phase retrieval (TADPOLE, STARFISH). These nonlinear interferometric techniques whether they employ Fourier transform spectroscopy or an intensity based detector, can overlap in the properties they provide but have overall been able to accurately characterize ultrafast optics over the past two decades. Other spatio-temporal characterization techniques have popped up each with their unique spin on characterizing a part of the beam. See examples of CROAK, HAMSTER, TERMITE, STRIPED FISH, INSIGHT.

While this list is not all encompassing, it shows the broad ranging methods to measure spectral contents, spatial widths, temporal widths, and other important properties. For a larger encompassing tutorial on characterization of pulses we direct you to S. W. Jolly et al. [Spencer W Jolly *et al* 2020 *J. Opt.* **22** 103501].

*Beam Deflection*

Beam deflection is a third order nonlinear measurement technique that measures both phase and amplitude changes caused by a modulated medium. It has been shown to produce equivalent results to the popular Z-scan method while retaining an advantage because it can measure temporal and spatial dynamics, polarization dependence, and non-degenerate excitation without experimental modifications.

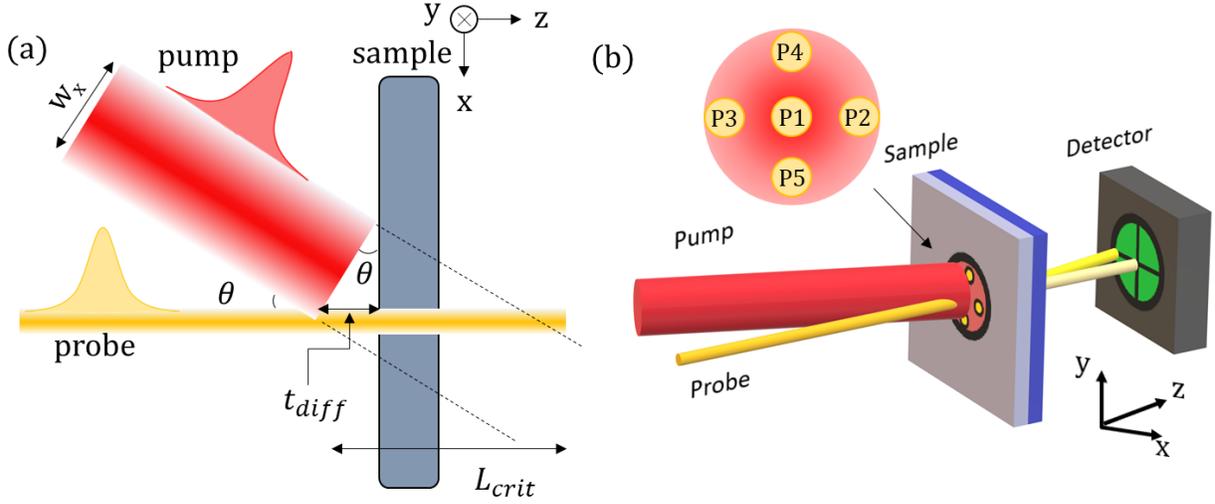

Figure: (a) XZ-plane of the pump beam arriving on a sample at an angle. The difference in time from the arrival of one side versus the other is denoted by $t_{diff}$. $L_{crit}$ defines the critical distance at which the two beams no longer overlapped. (b) Experimental setup of beam deflection showing a pump and probe beam temporally delayed and five possible positions of probe beams relative to a larger pump beam

In the ideal deflection case, the probe beam will deflect due to the refractive index gradient in the direction of higher index, e.g., at position 2 the probe beam will deflect to the left (resulting in a negative $\Delta E_x$) and at position 3 the probe beam will deflect to the right (resulting in a positive $\Delta E_x$) when monitoring the x-direction for a positive $n_2$ material. However, in positions 4 and 5, the probe beam will not deflect in the x-direction because there is no gradient present in the horizontal direction. The opposite is the case for the y-direction. A special point, position 1 in the center, should not deflect in either direction due to no spatial gradient being present in either the x- or y-direction. The probe is balanced equally by the radial distribution of energy. On the contrary, beam deflection can be found to have a gradient in the x-direction always in realistic setups.

*Energy Density in ENZ Thin Film*

Coordinate transformation for rotating a Gausssian beam in space with angle $\theta$.

$$\begin{bmatrix} a & b \\ b & c \end{bmatrix} = \begin{bmatrix} \frac{\cos^2(\theta)}{2w_x^2} + \frac{\sin^2(\theta)}{2w_y^2} & \frac{\sin(2\theta)}{4w_x^2} + \frac{\sin(2\theta)}{4w_y^2} \\ \frac{\sin(2\theta)}{4w_x^2} + \frac{\sin(2\theta)}{4w_y^2} & \frac{\sin^2(\theta)}{2w_x^2} + \frac{\cos^2(\theta)}{2w_y^2} \end{bmatrix} \quad (S1a)$$

From here, a rate equation can be used to determine the energy density in the sample with respect to both spatial and temporal coordinates. Nonlinearities are dependent upon the strength of the electric field and/or absorbed energy. Both are effectively a spatial energy density. The typical rate equation can be written in terms of energy density ($U$) and its general solution is shown by equations 4a-b:

$$\frac{dU(x,y,t)}{dt} = \alpha I(x,y,t) - \frac{U(x,y,t)}{\tau_R}, \quad (S2a)$$

$$U(x,y,t) = e^{-t/\tau_R} \int_{-\infty}^{t} \alpha I(x,y,t') e^{-t'/\tau_R} dt', \quad (S2b)$$

where $\alpha$ is the absorption and $\tau_R$ is the relaxation of the film. Applying the intensity profile of the beam from equation S2b and solving integral for the position of $z = 0$, results in the solution:

Solving the integral only:

$$\int_{-\infty}^{t} \alpha I(x,t') e^{-t'/\tau_R} dt'$$

$$= A\alpha \left(e^{-\frac{x^2}{2w_p^2}}\right) \int_{-\infty}^{t} \left(e^{-\frac{t'^2}{2\tau_p^2}}\right) e^{-t'/\tau_R} dt'$$

$$= A(x) \int_{-\infty}^{t} e^{-\frac{t'^2}{2\tau_p^2} - t'/\tau_R} dt'$$

Complete the square

$$= A(x) \int_{-\infty}^{t} e^{\left[\frac{\tau_p^2}{2\tau_R^2} - \left(\frac{t'}{\sqrt{2}\tau_p} + \frac{\tau_p}{\sqrt{2}\tau_R}\right)^2\right]} dt'$$

U-substitution

$$u = \frac{\tau_R t + \tau_p^2}{\sqrt{2}\tau_p \tau_R} \rightarrow du = \frac{1}{\sqrt{2}\tau_p}$$

$$= A(x) \int_{-\infty}^{t} \left(\frac{\sqrt{\pi}}{\sqrt{2}} \tau_p e^{\frac{\tau_p^2}{2\tau_R^2}}\right) \left(\frac{2e^{-u^2}}{\sqrt{\pi}}\right) du$$

Substituting in for erf(u)

$$\int \left(\frac{2e^{-u^2}}{\sqrt{\pi}}\right) du = \text{erf}(u)$$

$$= A(x) \int_{-\infty}^{t} \left(\frac{\sqrt{\pi}}{\sqrt{2}} \tau_p e^{\frac{\tau_p^2}{2\tau_R^2}}\right) \left(\frac{2e^{-u^2}}{\sqrt{\pi}}\right) du$$

$$= A(x)\left(\frac{\sqrt{\pi}}{\sqrt{2}}\tau_p e^{\frac{\tau_p^2}{2\tau_R^2}}\right)\text{erf}(u)$$

Substitute back u in and evaluate

$$= A(x)\frac{\sqrt{\pi}}{\sqrt{2}}\tau_p e^{\frac{\tau_p^2}{2\tau_R^2}}\text{erf}\left(\frac{\tau_R t' + \tau_p^2}{\sqrt{2}\tau_p\tau_R}\right)\Bigg|_{-\infty}^{t}$$

$$= A(x)\frac{\sqrt{\pi}}{\sqrt{2}}\tau_p e^{\frac{\tau_p^2}{2\tau_R^2}}\left[\text{erf}\left(\frac{\tau_R t + \tau_p^2}{\sqrt{2}\tau_p\tau_R}\right) + 1\right]$$

Plug back into rate equation general solution

$$U(x,y,t) = e^{-t/\tau_R}\int_{-\infty}^{t}\alpha I(x,y,t')e^{-t'/\tau_R}dt'$$

$$U(x,y,t) = \left(A(x,y)\frac{\sqrt{\pi}}{\sqrt{2}}\tau_p e^{\frac{\tau_p^2}{2\tau_R^2}}\left[\text{erf}\left(\frac{\tau_R t + \tau_p^2}{\sqrt{2}\tau_p\tau_R}\right) + 1\right]\right)e^{-t/\tau_R}$$

$$U(x,y,t) = \left(A(x,y)\frac{\sqrt{\pi}}{\sqrt{2}}\tau_p e^{\frac{\tau_p^2}{2\tau_R^2}}\left[\text{erf}\left(\frac{\tau_R t + \tau_p^2}{\sqrt{2}\tau_p\tau_R}\right) + 1\right]\right)e^{-\frac{t}{\tau_R}} \quad (S3)$$

For the off-axis case:

$$U(x,t) = e^{-t/\tau_R}\int_{-\infty}^{t}\alpha I(x,t')e^{-t'/\tau_R}dt'$$

$$I(x,t) = Ae^{-a(x-x_0)^2}e^{-2b(x-x_0)(z-z_0)}e^{-c(z-z_0)^2}$$

$$x_0 = v_{g_x}t \rightarrow v_{g_x} = v_g\sin(\theta)$$

$$z_0 = v_{g_z}t \rightarrow v_{g_z} = v_g\cos(\theta)$$

Evaluating at z = 0 (sample surface)

$$I(x,t) = Ae^{-a(x-x_0)^2}e^{-2b(x-x_0)(-z_0)}e^{-c(-z_0)^2}$$

Solving with completing the square and u-sub again results in:

$$U(x,y,t) = e^{-\frac{t(x)}{\tau_R}} \left[ \frac{-\sqrt{\pi} e^{-\frac{4v\tau_R x(acv\tau_R x - b^2 v\tau_R x + b + a) - 1}{4(c+2b+a)v^2\tau_R^2}} \cdot \left( \text{erf}\left( \frac{2v\tau_R(bx + ax - ctv - 2btv - atv) - 1}{2\sqrt{c+2b+a}v\tau_R} \right) - 1 \right)}{2\sqrt{c+2b+a}v} \right] \quad (S4)$$

This results in varying energy distributions and slopes dependent upon probe position, and excitation incident angle as shown below.

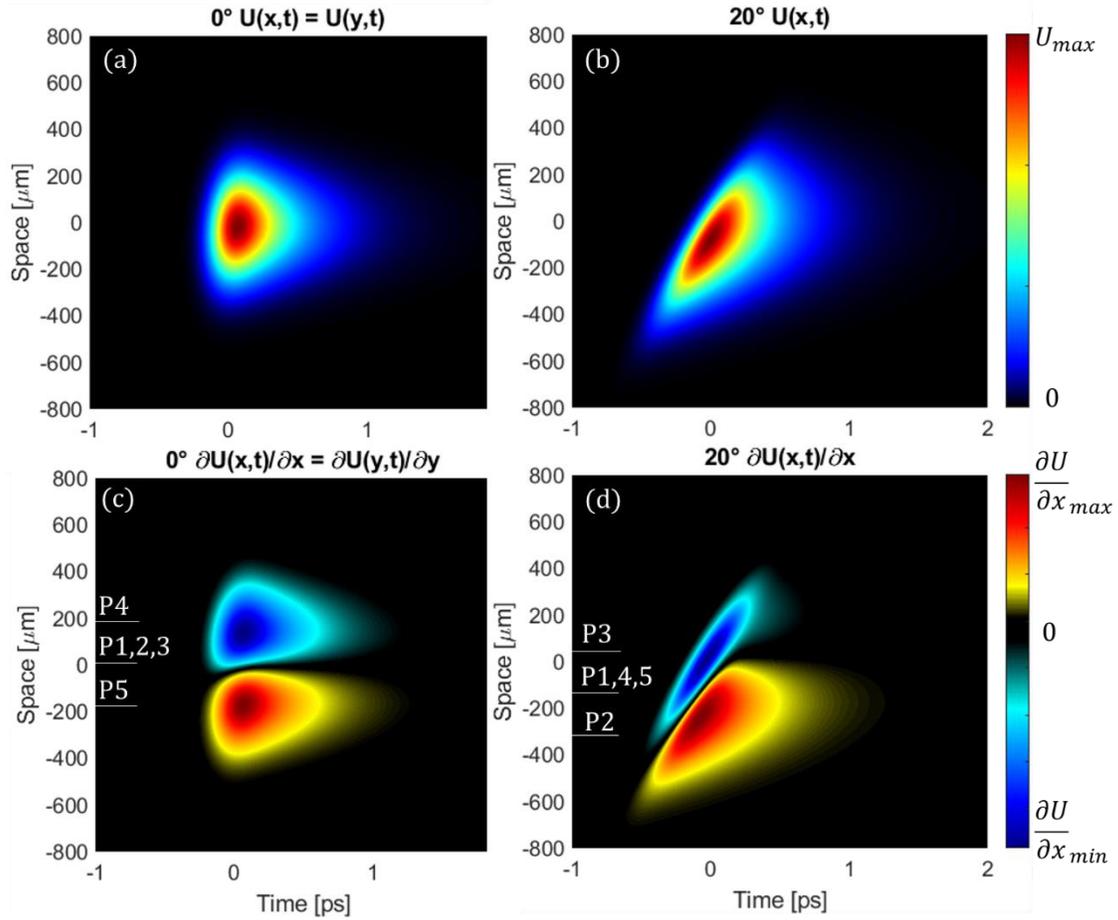

Figure: (a) Distribution of energy in space and time for an excitation beam at normal incidence (0 degrees) at the sample surface. In the case of normal incidence U(x,t) and U(y,t) are equal. (b) Distribution of energy in space and time for an excitation beam at 20 degrees. (c) Slope of surface plot (a) for normal incidence, the slopes are equivalent at all positions. (d) Slopes ∂U(x,t)/∂x of surface plot (b) at 20 degrees, where there is a bipolar response because of the distribution of the pulse in space and time.

## Supplemental 2: Experimental details

Solstice Ace 800nm 1kHz 100fs pumps dual OPAs with tunable wavelengths between 1200-1600 nm. The experiments were done in a non-degenerate combination with a pump beam at 1400 nm and a probe beam at 1300 nm (~35 nm FWHM measured in a spectrometer) with cross-polarization. The pulse width of the Solstice was measured in a FROG system with a width of ~97 fs. The excitation and probe beams were measured at each angle using a BeamOn WSR CCD beam profiler. Below are images of the beams and cross-sectional sizes. The probe beam is ~170μm in size (1/e² diameter) and the excitation beam is ~600μm in size (1/e² diameter).

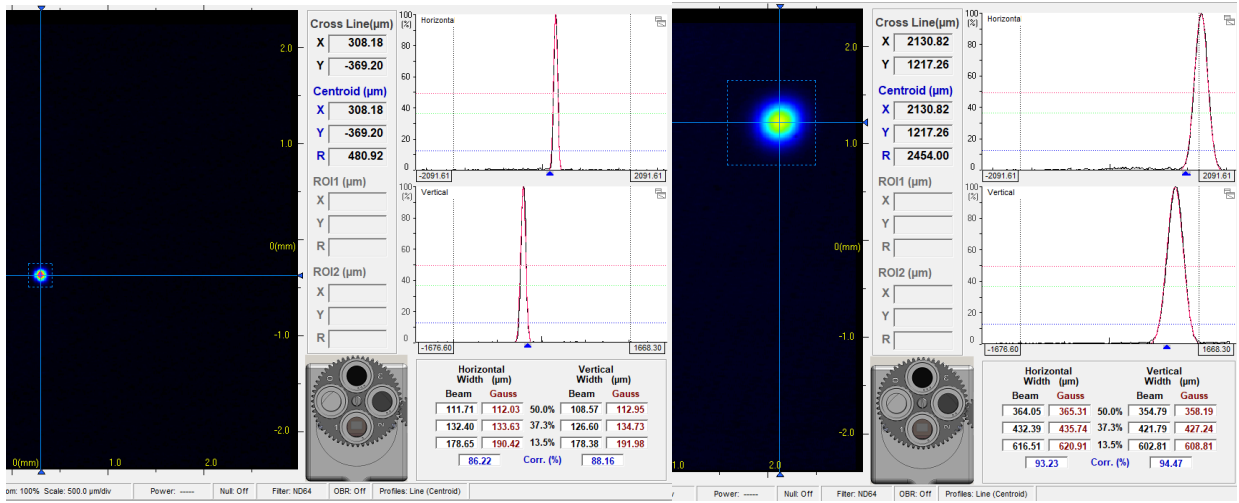

The intensity of the excitation beam is 56 GWcm$^{-2}$. Measurements were done on a 240 nm Al:ZnO film whose permittivity can be seen below: The crossover point is ~1560 nm with a loss of $\varepsilon'' = 0.46$.

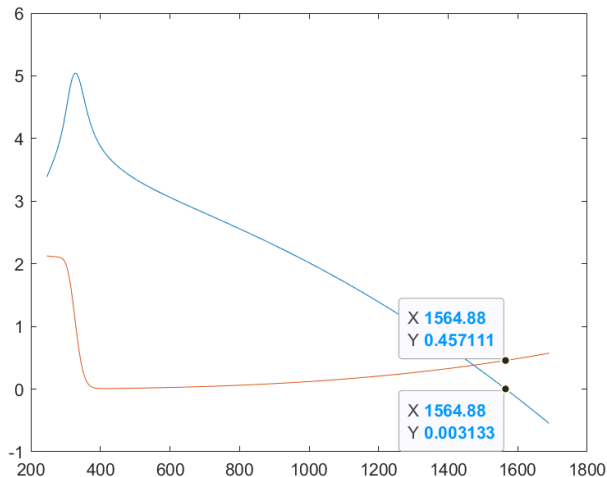

*Space-Time Knife Edge in Fused Silica*

If we begin with one of the simplest examples of an intensity dependent refractive index, we can build up our model of off-axis excitation and see how it applies to geometric parameters. Fused silica (FS) has an instantaneous Kerr nonlinearity $n = n_0 + n_2 I$. Completing the same five-point measurements that were done in Al:ZnO gives the following (with an increased intensity of 175 GWcm$^{-2}$):

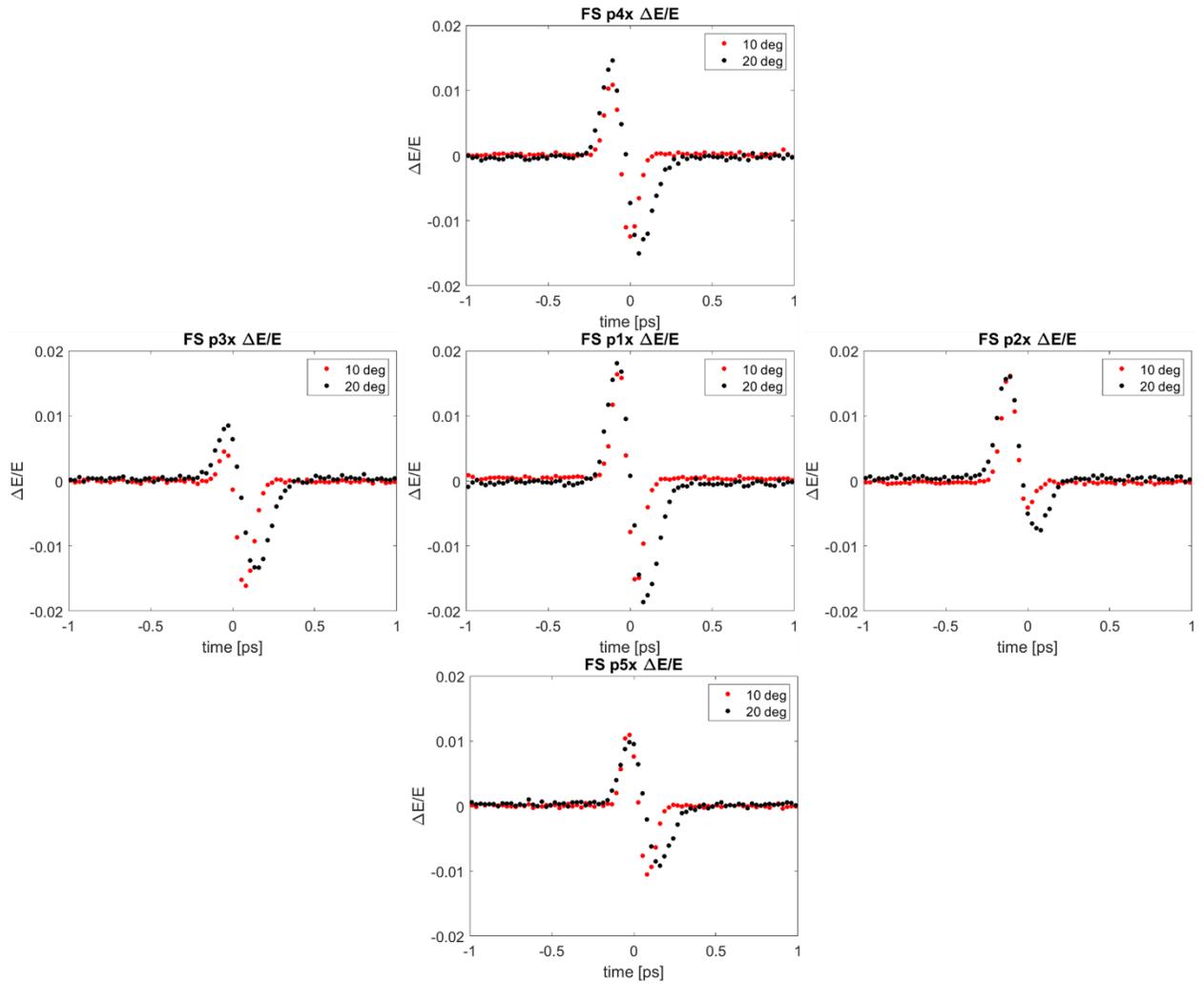

The sweeping motion present in the x-direction is shown in all cases. The important distinction between this and the ENZ film is the thickness of the FS film of 500μm. This allows for walk-off in the larger angle cases, although the angular-induced gradient from the arrival time of the beam is larger. So, there are two contradicting parameters determining the output signal.

Additionally, this signal is much weaker than the ENZ film's signal due to the differences in $n_2$ being so stark. The side effect is that the FS film has no relaxation time, corresponding much closer to the output of an intensity autocorrelation function. These signals can be modeled using equation 5, with a near instantaneous relaxation time.

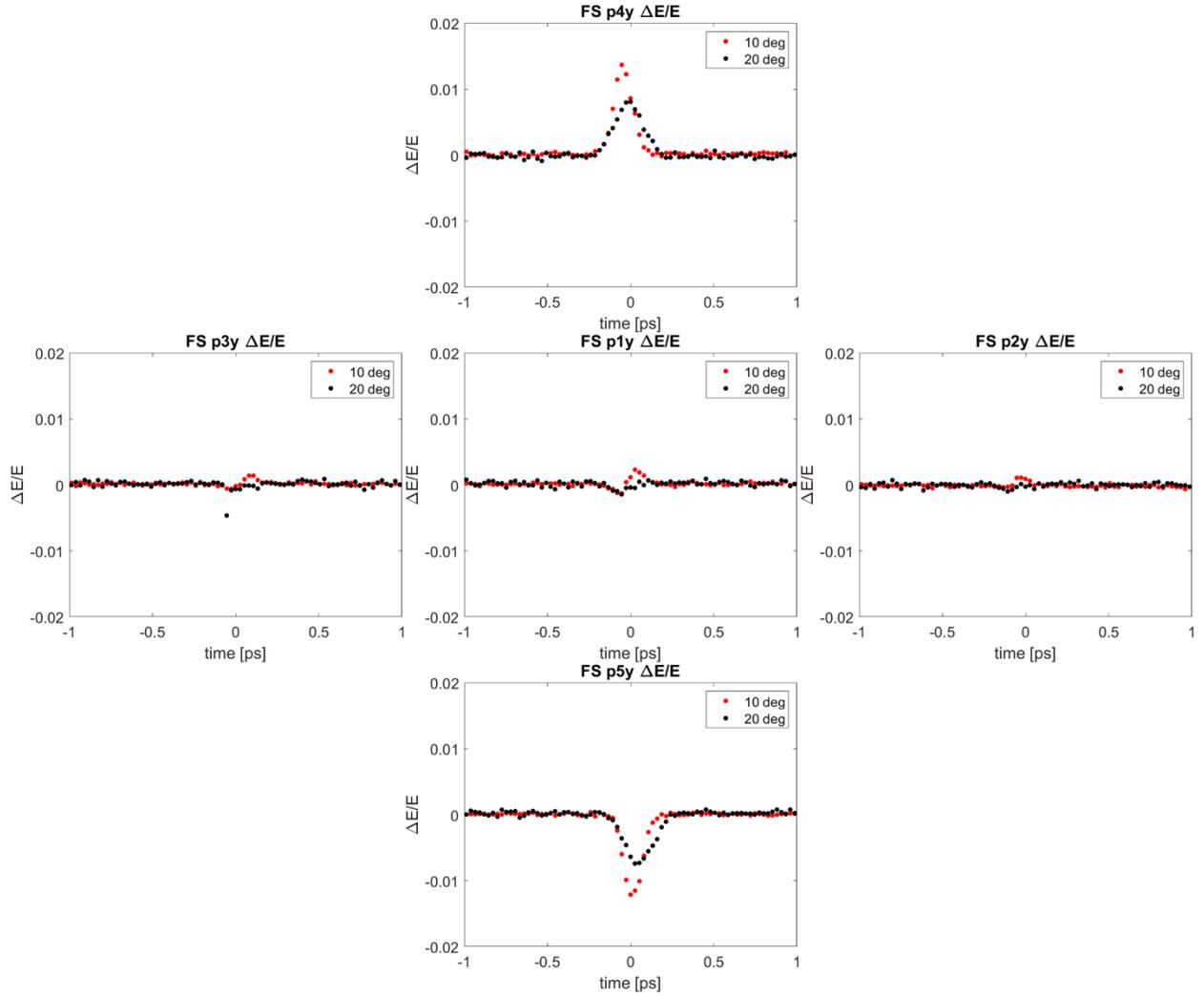

In the vertical direction, the signals have the same signatures as the ENZ film does where positions P1-P3 show no deflection because there is no vertical gradient. In positions P4 and P5, the ideal nonlinearity is shown. The big difference is that at a higher angle, the nonlinearity drops significantly due to the shorter interaction times between the beams. The parameter L$_{crit}$ reduces the overlap and thus we have a smaller BD signal. In the two extreme cases, if the angle is zero (the two beams are colinear) the critical thickness is infinite, and if the angle is 90 degrees (two beams are perpendicular), the critical thickness is the width of the excitation beam (that is strong enough to modulate $n$, typically the FWHM). In such cases where $L_{sample} > L_{crit}$, the two beams could walk off and decrease the nonlinear interaction according to the percentage of distance where the beams are overlapped within the sample compared to the entire thickness of the sample.

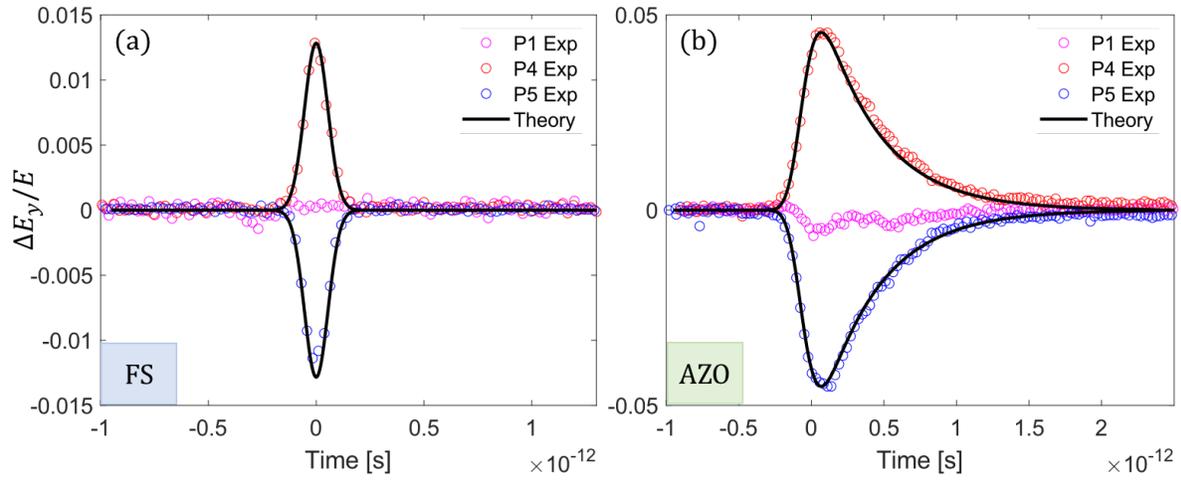

A model of the two ideal signals overlapped with experimental data is shown here. This is taken at positions P1, P4, and P5, and modeled by equation 5 at zero degrees angle of incidence. This proves that the model works for both normal and off-axis excitation.